%% file: main.tex
\documentclass[a4paper,11pt]{article}
\usepackage{pos}
\usepackage{xspace}

\title{Recent results on searches with boosted Higgs bosons at CMS}

\author*{Farouk Mokhtar}
\onbehalf{on behalf of the CMS Collaboration}

\affiliation[]{Department of Physics \\
University of California San Diego \\
9500 Gilman Drive \\
La Jolla, CA 92093
}

\emailAdd{fmokhtar@ucsd.edu}

\abstract{
The study of boosted Higgs bosons at the LHC provides a unique window to probe Higgs boson couplings at high energy scales and search for signs of physics beyond the standard model.
In these proceedings, we present recent results on boosted Higgs boson searches at the CMS experiment, highlighting innovative reconstruction and tagging techniques that enhance sensitivity in this challenging regime.
}

\FullConference{The Thirteenth Annual Large Hadron Collider Physics (LHCP2025)\\
5-9 May 2025\\
Taipei\\}

\input{commands}

\begin{document}
\maketitle

\section{Introduction}

The discovery of the Higgs boson (\PH) by ATLAS~\cite{ATLAS:2012yve} and CMS~\cite{CMS:2012qbp} at the LHC completed the standard model (SM) and opened new avenues to study electroweak symmetry breaking.
Since then, a core focus has been precision measurements of \PH properties and couplings to test the SM and search for beyond the SM (BSM) physics.
At high transverse momentum (\pt), \PH production probes higher energy scales and enhances sensitivity to anomalous couplings.
The boosted regime, while challenging due to small SM cross sections, results in decay products that merge into large-radius jets, enabling jet substructure and machine learning (ML)-based taggers to separate signal from background.
In these proceedings, we summarize innovative methods for boosted \PH topologies and recent results from the CMS experiment~\cite{Chatrchyan:2008zzk} at the LHC.

\section{ML-based jet tagging for boosted \PH topologies}

Notable advances in CMS include the development of state-of-the-art large-radius jet taggers such as ParticleNet-MD~\cite{Qu:2019gqs,CMS-PAS-BTV-22-001}, with a graph neural network (GNN) architecture, and Global Particle Transformer (GloParT)~\cite{Qu:2022mxj,CMS-PAS-HIG-23-012}, with an attention-based transformer architecture.
These architectures efficiently exploit particle-level pairwise features within jets to infer their origin.
ParticleNet-MD classifies jets into 8 categories, specifically targeting $\PX(\Pq\Pq)$ decays, while GloParT targets a broader set of 37 classes, including $\PX(\PV\PV)$ decays and top quark-antiquark pairs (\ttbar).

\begin{figure}[ht]
    \centering
    \includegraphics[width=0.305\linewidth]{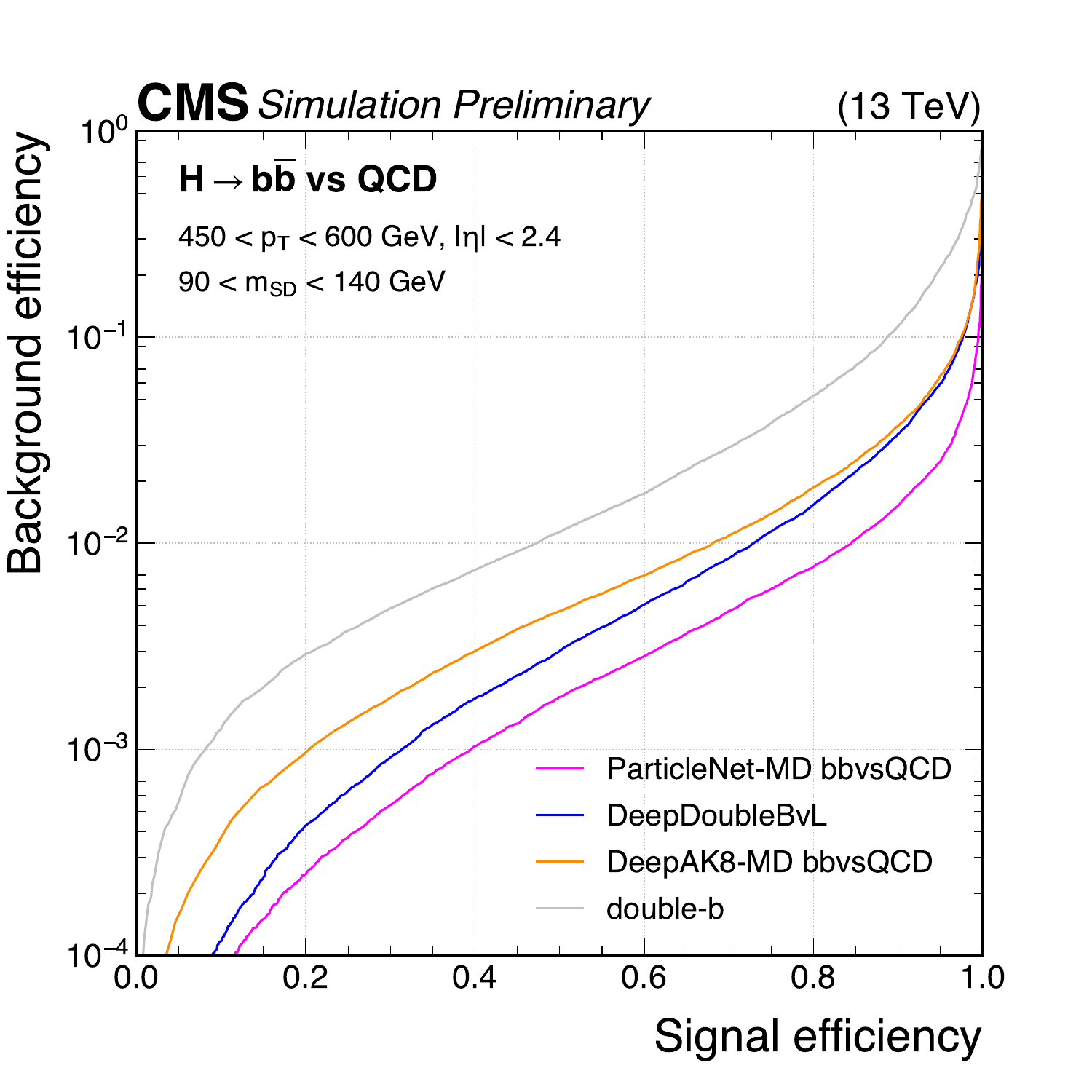}
    \includegraphics[width=0.305\linewidth]{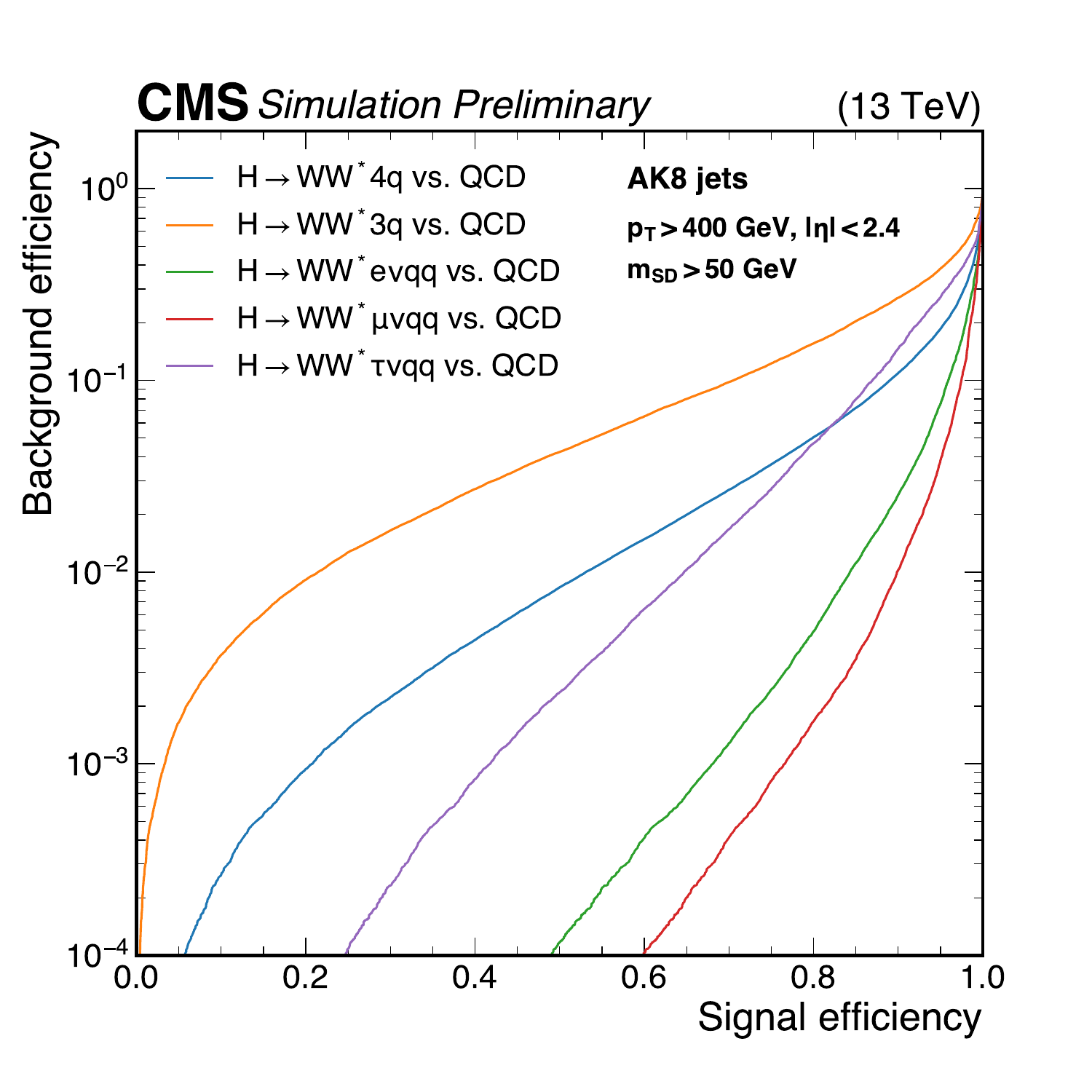}
    \includegraphics[width=0.305\linewidth]{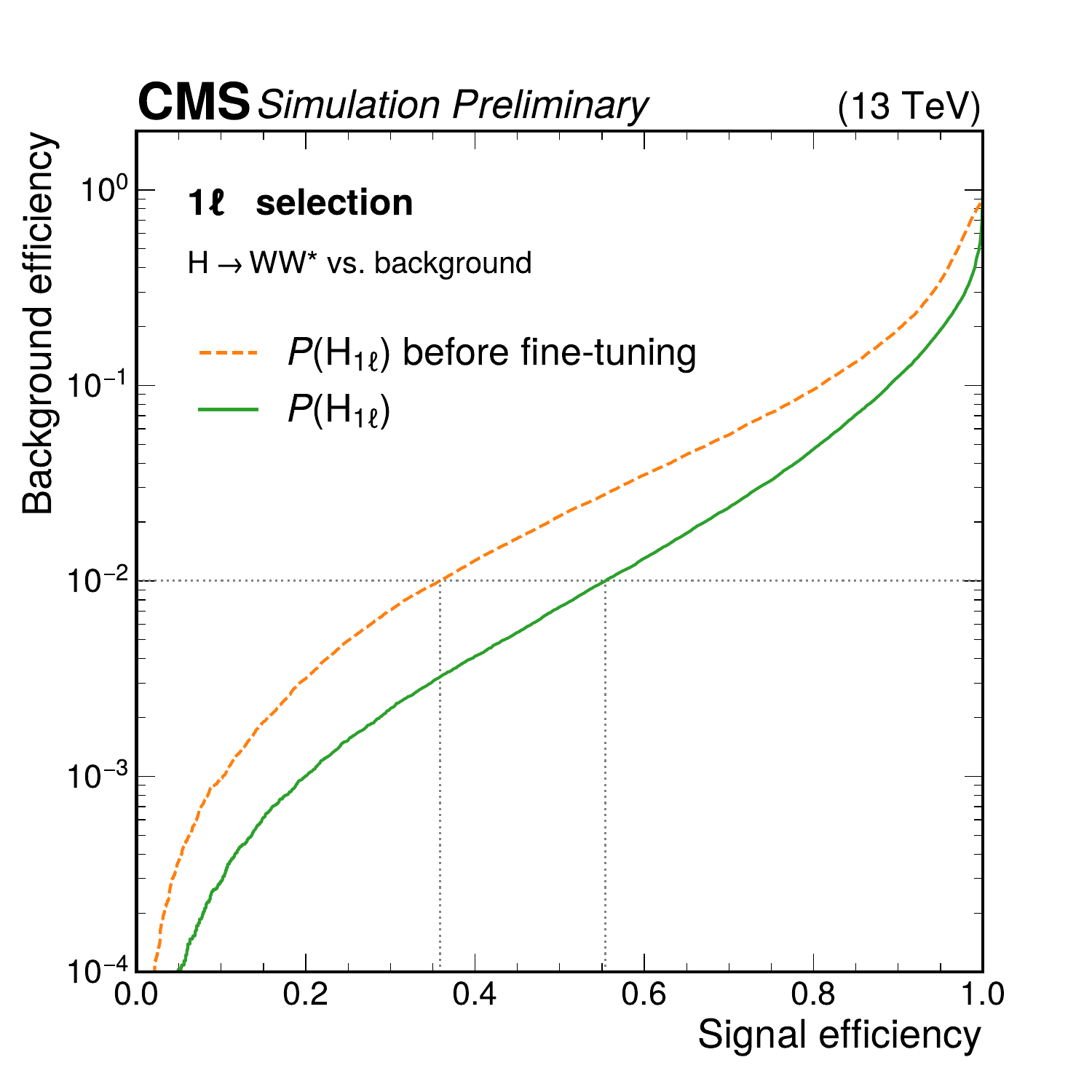}
    \caption{
    Left: Performance of ParticleNet-MD for \hbb tagging~\cite{CMS-PAS-BTV-22-001}. Middle: GloParT for \hww tagging~\cite{CMS-PAS-HIG-24-008}. 
    Right: Fine-tuned version of GloParT targeting $1\ell$ final state topologies~\cite{CMS-PAS-HIG-24-008}.
    }
    \label{fig:roc}
\end{figure}

The tagger performance is quantified via receiver operating characteristic (ROC) curves.
Figure~\ref{fig:roc} (left) shows the performance of ParticleNet-MD for \hbb tagging, achieving strong QCD multijet background rejection and outperforming DeepAK8-MD~\cite{JME-18-002}, DeepDoubleX~\cite{CMS-DP-2022-041}, and the double-b tagger~\cite{Sirunyan:2017ezt}. 
Figure~\ref{fig:roc} (middle) highlights GloParT's performance for \hww tagging, with results shown split by final state.
Moreover, large ML models like GloParT can be fine-tuned on specific signal topologies using transfer learning, where the model is further trained on a smaller, targeted dataset. 
In the one-lepton ($1\ell$) channel of the boosted \hww analysis~\cite{CMS-PAS-HIG-24-008}, a fine-tuned GloParT model ($P(\PH_{1\ell})$) was developed to capture the signature of isolated leptons overlapping with large-radius jets that was not well represented in the original training, leading to a nearly 70\% increase in expected significance for that channel.
Figure~\ref{fig:roc} (right) shows the performance of $P(\PH_{1\ell})$ against a background consisting of \ttbar, \wjets, and QCD multijets before and after fine-tuning.

\section{Boosted \texorpdfstring{\hww}{H to WW}}

In the analysis in Ref.~\cite{CMS-PAS-HIG-24-008}, \PH boson decay products are merged into a single high-\pt large-radius jet.
Two categories are defined: the 0$\ell$ channel (no isolated leptons, QCD dominated, \PH candidate identified using GloParT) and the 1$\ell$ channel (one isolated $e$/$\mu$, dominated by \wjets and \ttbar, \PH candidate identified using $P(\mathrm{H_{1\ell}})$).
The 1$\ell$ category is further split by production mode, gluon fusion (ggF) and vector boson fusion (VBF).
The 0$\ell$ category is categorized by a GloParT tagger cut and a selection on missing transverse momentum: $\met >20$~\GeV.
The QCD multijets background in 0$\ell$ is estimated using extrapolation from data in control region.
Finally, since GloParT is trained in Monte Carlo (MC) simulation, the Lund jet plane reweighting technique~\cite{CMS:2025yce,CMS-DP-2023-046} is used to calibrate it in data.
This method provides a generalized calibration procedure that can be applied to jets with any number of prongs, in particular where no SM analogue for the signal exists. 

The signal is extracted from a binned maximum likelihood fit to the H candidate softdrop mass (\msd) distribution, where \msd is the jet mass after softdrop grooming~\cite{Larkoski:2014wba}.
The results are shown in Fig.~\ref{fig:hww} (left).
The observed significance for the combined result is 0, while the expected significance is $1.76\sigma$.
This is the first dedicated study of boosted \hww decays in CMS.

\section{Boosted \texorpdfstring{\hhbbvv}{HH to bbVV}}

\begin{figure}[ht]
    \centering
    \includegraphics[width=0.62\linewidth]{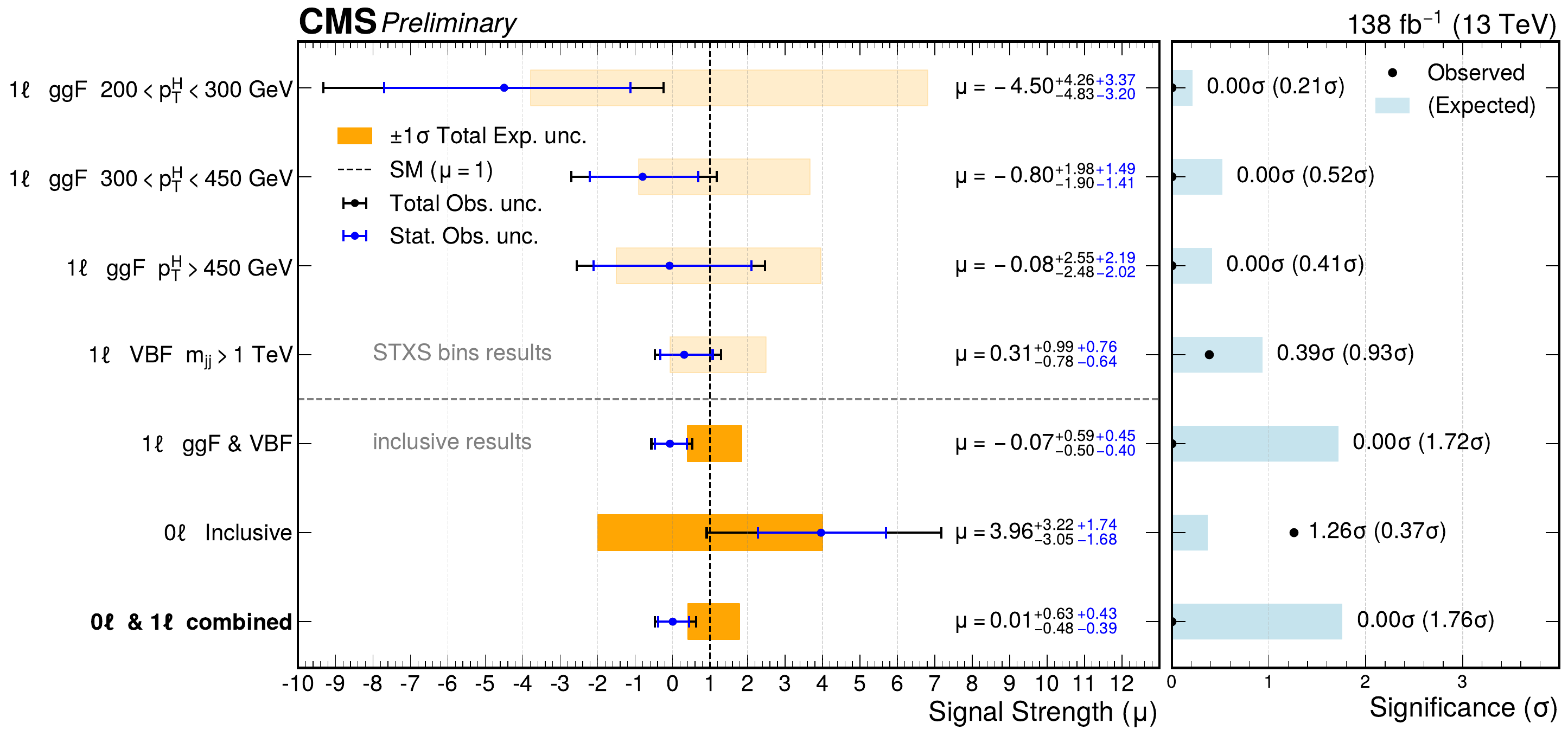}
    \includegraphics[width=0.37\linewidth]{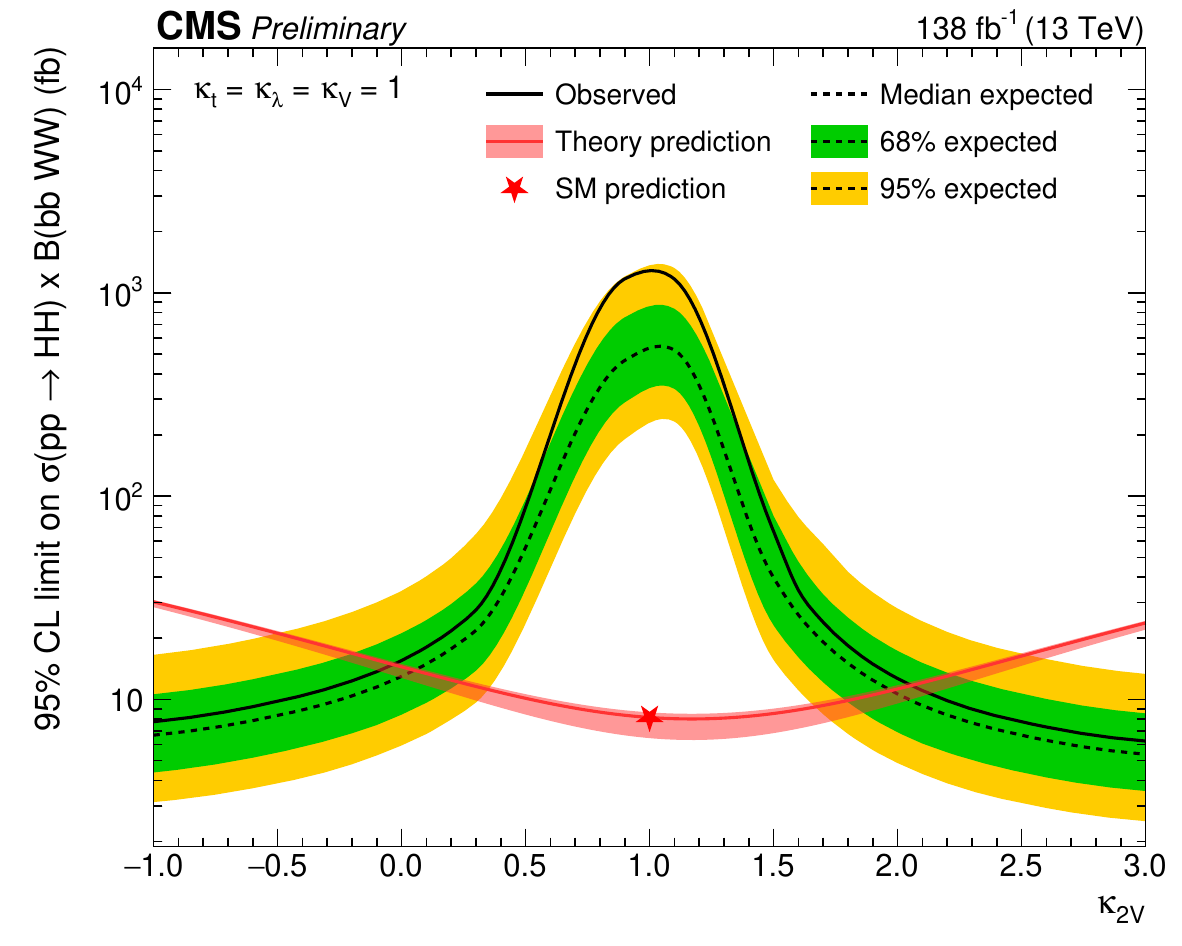}    
    \caption{
    Left: Fitted signal strength and significance for the \hww analysis~\cite{CMS-PAS-HIG-24-008}.
    Right: Upper limits on the inclusive $\PH\PH$ production cross section as a function of $\kappa_{2\PV}$~\cite{CMS-PAS-HIG-23-012}.
    }
    \label{fig:hww}
\end{figure}

In the analysis in Ref.~\cite{CMS-PAS-HIG-23-012}, large-radius jets are identified using ParticleNet-MD for \hbb and GloParT for \hww.
The mass observable for signal extraction is the H candidate mass after a regression is applied to improve the mass resolution.
The signal region is categorized by production mode (ggF/VBF), and the QCD multijets background estimation is performed using extrapolation from data in control region.
Similar to the boosted \hww analysis, GloParT is calibrated using Lund jet plane reweighting.
The results are shown in Fig.~\ref{fig:hww} (right) and delivers the second-strongest CMS constraints on the the quartic
$\PV\PV\PH\PH$ coupling modifier $\kappa_{2\PV}$~\cite{CMS:2022dwd}, the most sensitive one being boosted \hhbbbb~\cite{CMS-PAS-B2G-22-003}.

\section{Boosted \texorpdfstring{\vhbb}{VH to bb}}

At high \pt, the \vh production mode with the \hbb final state has the largest branching fraction and SM cross section.
This analysis~\cite{CMS-PAS-HIG-24-017} is motivated by the VBF excess observed in Ref.~\cite{HIG-21-020}, which could be a sign of an enhanced coupling to vector bosons.
Events are selected using jet triggers, a lepton veto, and $\ptmiss<140\GeV$ to suppress \ttbar background. 
ParticleNet-MD identifies both \vqq and \hbb jets.
Backgrounds are estimated from MC, except QCD multijets, which is estimated using extrapolation from data in control region.

The \PV candidate jet \msd distribution is used for signal extraction (Fig.~\ref{fig:vh}, left).
The fitted signal strengths for \vh and \vz production, denoted $\mu_{\vh}$ and $\mu_{\vz}$ respectively, are extracted following Ref.~\cite{CMS:2014fzn}, and a combined fit over the full data set is performed for the final result. 
The ggF, VBF, and $\mathrm{t\bar{t}H}$ processes are normalized and fixed to their SM predictions. 
The combined $\mu_{\vh}$ is measured (expected) to be $\mu_{\vh} = 0.72^{+0.75}_{-0.71} \; (1.00^{+0.67}_{-0.62})$, 
and the combined $\mu_{\vz}$ is measured (expected) to be $\mu_{\vz} = 0.09\pm0.63 \; (1.00\pm0.58)$.
This corresponds to an observed (expected) SM significance of $1.00(1.64)\sigma$ for $\PV(\qq)\PH(\bb)$ and $0.15(1.76)\sigma$ for $\PV(\qq)\PZ(\bb)$, marking the first dedicated CMS search for boosted \vhbb.

\section{\texorpdfstring{$\PH\to\PZ\PZ/\bb+\gamma$}{H to ZZ/bb + gamma}}

\begin{figure}[ht]
    \centering
    \includegraphics[width=0.315\linewidth]{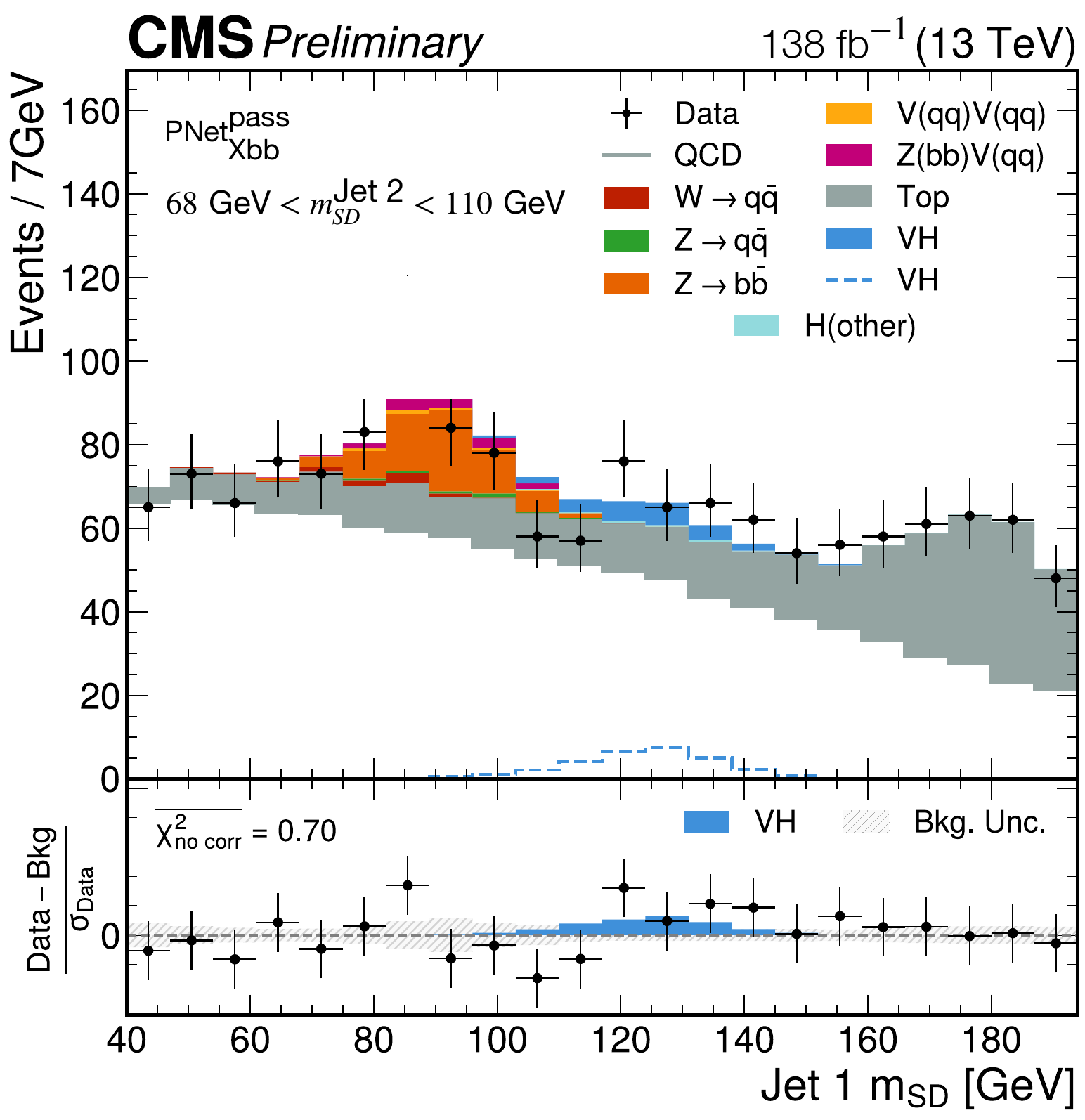}
    \includegraphics[width=0.315\linewidth]{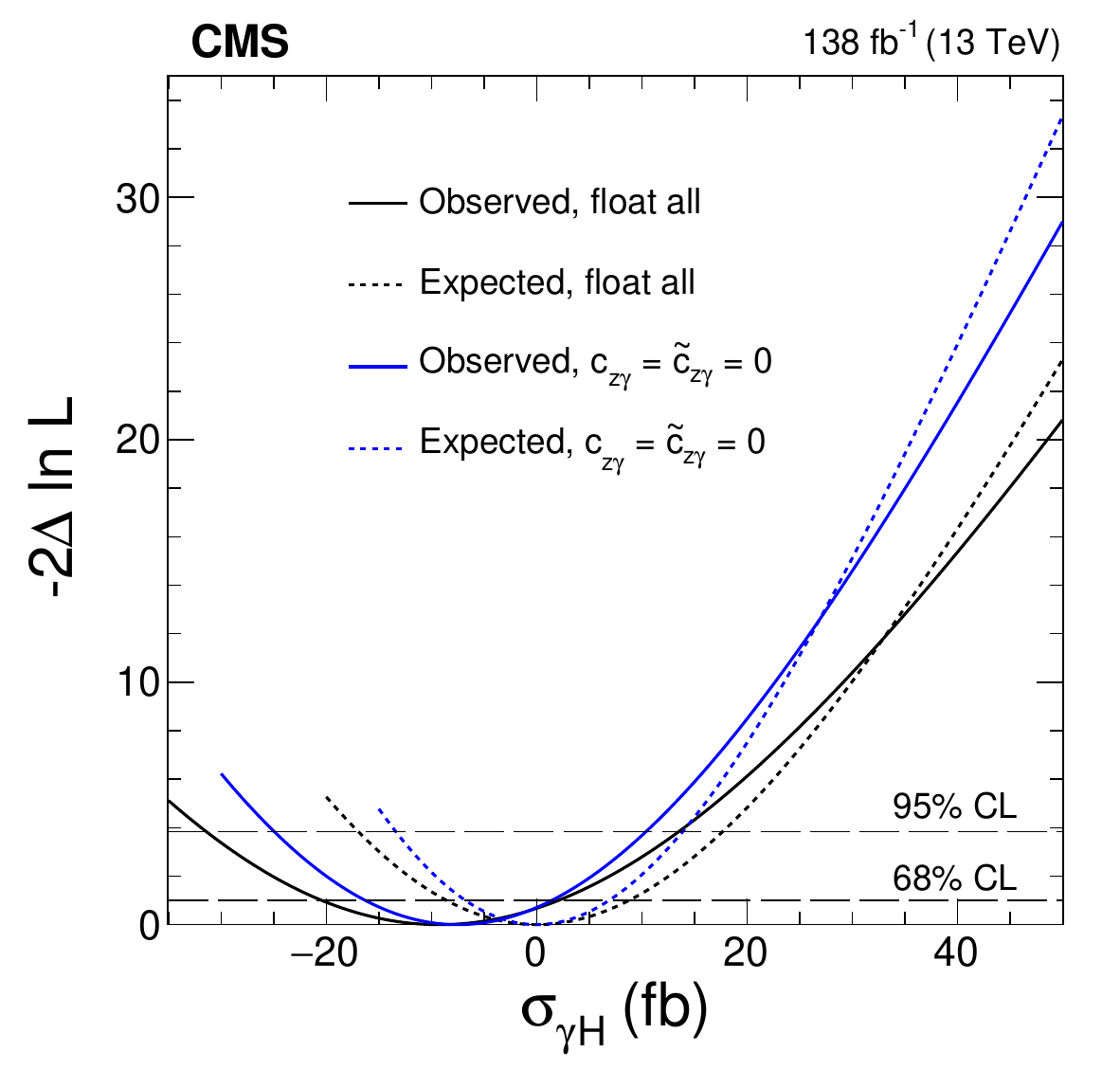}
    \includegraphics[width=0.315\linewidth]{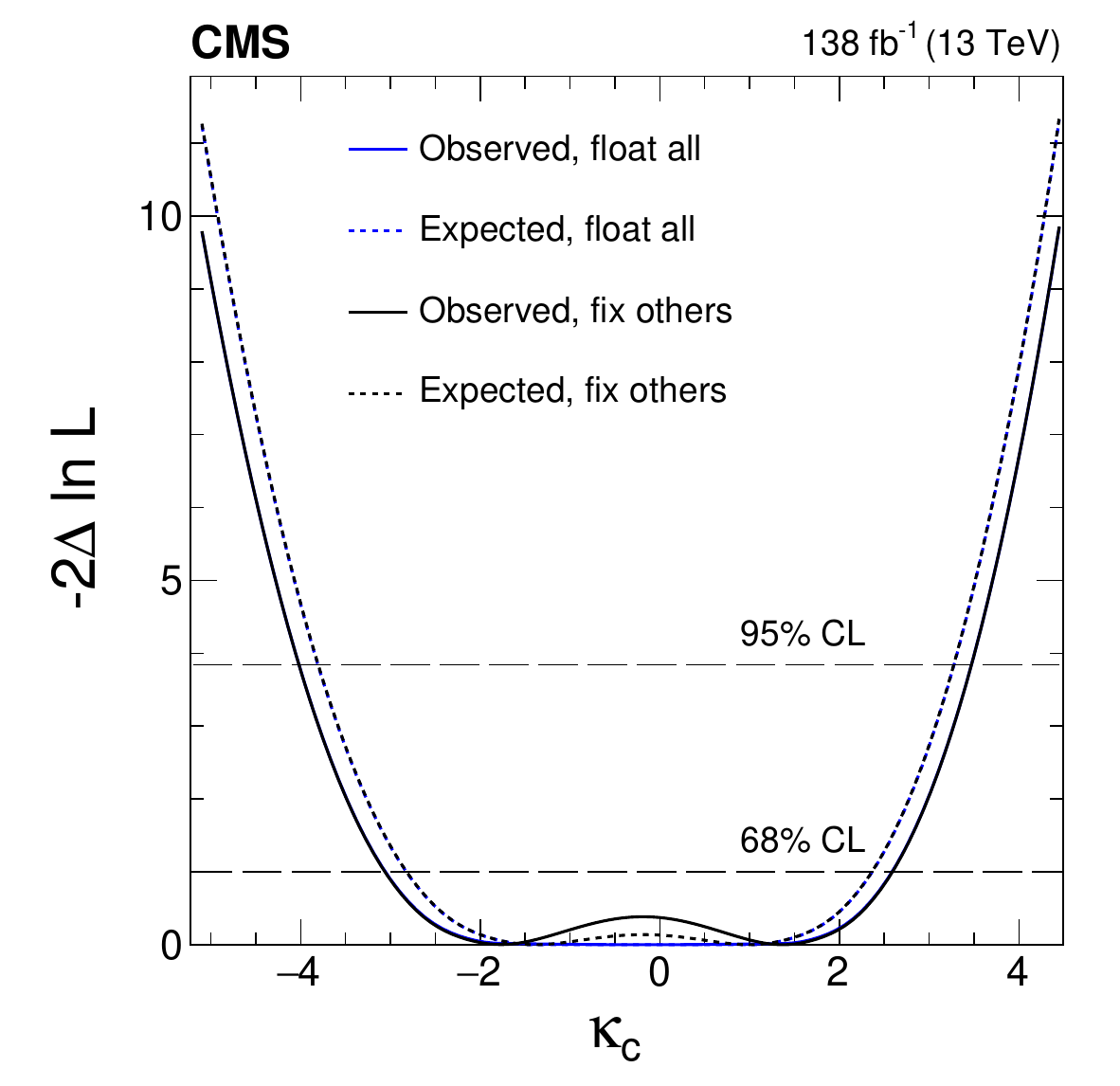}
    \caption{
    Left: Post-fit H boson candidate \msd distribution in the signal region, summed over all data-taking periods~\cite{CMS-PAS-HIG-24-017}.
    Middle: Constraints on $\sigma_{\gamma \mathrm{H}}$ using \hbb and \hllll~\cite{Chekhovsky:2923874}.
    Right: Constraints on the Yukawa couplings of light quarks using \hllll~\cite{Chekhovsky:2923874}.
    }
    \label{fig:vh}
\end{figure}

Finally, the first $\gamma$H results at the LHC are presented~\cite{Chekhovsky:2923874}.
Although the SM cross section is small, the observation of such a signal would provide clear evidence of BSM physics. 
The analysis is split into two categories: (i) $\gamma$H $\to$ ZZ $\to$ 4$\ell$, with four isolated leptons, and (ii) \ghbb, where a single large-radius jet is identified using ParticleNet-MD and the QCD multijet background is estimated using extrapolation from data in control region.
The results are limited by statistical precision.
Constraints on $\sigma_{\gamma \mathrm{H}}$ are set using both \hbb and \hllll decays, and on the Yukawa couplings of light quarks using \hllll, as shown in Fig.~\ref{fig:vh}.

\section{Summary and outlook}

The CMS collaboration has extensively studied single Higgs boson (\PH) and Higgs boson pair ($\PH\PH$) production in boosted topologies across production modes, leveraging novel large-radius deep learning jet taggers.
High-Luminosity LHC (HL-LHC) data will be essential to enhance statistical power and probe these processes further. 
Exciting upcoming developments include the first use of ParticleNet $\bb$ triggers and improved GloParT taggers, promising significant sensitivity gains.

\bibliographystyle{cms_unsrt}
\bibliography{references}


\end{document}

%% file: commands.tex
\newcommand{\msd}{\ensuremath{m_{\mathrm{SD}}}\xspace}
\newcommand{\pt}{\ensuremath{p_{\mathrm{T}}}\xspace}
\newcommand{\ptmomentum}{\ensuremath{p_{\mathrm{T}}}\xspace}
\newcommand{\ptvecmiss}{\ensuremath{{\vec p}_{\mathrm{T}}^{\kern1pt\text{miss}}}\xspace}
\newcommand{\ptmiss}{\ensuremath{\ptmomentum^\text{miss}}\xspace}
\newcommand{\met}{\ptmiss}

\newcommand{\akfourchs}[1]{AK4-CHS\xspace}
\newcommand{\akfourpuppi}[1]{AK4-PUPPI\xspace}
\newcommand{\GeV}{\ensuremath{\,\text{Ge\hspace{-.08em}V}}\xspace}

\newcommand{\PQb}{\ensuremath{\mathrm{b}}\xspace}
\newcommand{\PZ}{\ensuremath{\mathrm{Z}}\xspace}
\newcommand{\PH}{\ensuremath{\mathrm{H}}\xspace}
\newcommand{\PV}{\ensuremath{\mathrm{V}}\xspace}
\newcommand{\Pq}{\ensuremath{\mathrm{q}}\xspace}
\newcommand{\PW}{\ensuremath{\mathrm{W}}\xspace}
\newcommand{\PX}{\ensuremath{\mathrm{X}}\xspace}

\newcommand{\ttbar}{\ensuremath{\mathrm{t\overline{t}}}\xspace}
\newcommand{\qq}{\ensuremath{\mathrm{q\overline{q}}}\xspace}
\newcommand{\bb}{\ensuremath{\mathrm{b\overline{b}}}\xspace}

\newcommand{\vh}{\ensuremath{\PV\PH}\xspace}
\newcommand{\vz}{\ensuremath{\PV\PZ}\xspace}

\newcommand{\hww}{\ensuremath{\PH \to \PW\PW}\xspace}
\newcommand{\hbb}{\ensuremath{\PH \to \bb}\xspace}
\newcommand{\vqq}{\ensuremath{\PV \to \qq}\xspace}

\newcommand{\wjets}{\ensuremath{\PW\text{+jets}}\xspace}
\newcommand{\hhbbvv}{\ensuremath{\PH\PH \to \bb\PV\PV}\xspace}
\newcommand{\hhbbbb}{\ensuremath{\PH\PH \to 4\PQb}\xspace}
\newcommand{\vhbb}{\ensuremath{\PV\PH \to \bb}\xspace}
\newcommand{\ghbb}{\ensuremath{\gamma \PH \to \bb}\xspace}

\newcommand{\hllll}{\ensuremath{\PH \to 4\ell}\xspace}